\documentclass{webofc}
\usepackage{subfig}
\usepackage{lipsum}
\usepackage[varg]{txfonts}   
\begin{document}
\title{(Anti)nucleosynthesis in heavy-ion collisions and (anti)nuclei as "baryonmeter" of the collision}
%
%

\author{\firstname{Mario} \lastname{Ciacco}\inst{1}\fnsep\thanks{\email{mario.ciacco@cern.ch}} for the ALICE Collaboration
}

\institute{Dipartimento DISAT del Politecnico and Sezione INFN, Turin, Italy}

\abstract{%
The production mechanism of light (anti)nuclei in heavy-ion collisions has been extensively studied experimentally and theoretically. Two competing (anti)nucleosynthesis models are typically used to describe light (anti)nuclei yields and their ratios to other hadrons in heavy-ion collisions: the statistical hadronization model (SHM) and the nucleon coalescence model. The possibility to distinguish these phenomenological models calls for new experimental observables.

Given their large baryon number, light (anti)nuclei have a high sensitivity to the baryon chemical potential ($\mu_{\rm B}$) of the system created in the collision. 

In this talk, the first measurement of event-by-event antideuteron number fluctuations in heavy-ion collisions is presented and compared with expectations of the SHM and coalescence model. In addition, the antinuclei-to-nuclei ratios are used to obtain a measurement of $\mu_{\rm B}$ in heavy-ion collisions with unprecedented precision.
}
\maketitle

\section{Introduction}
\label{intro}
(Anti)nucleosynthesis is one of the open questions of heavy-ion physics. While it is yet to be fully understood how loosely bound objects such as light (anti)nuclei can survive the extreme conditions reached in high-energy nuclear collisions, the general properties of their production can be described by two different phenomenological approaches, namely the Statistical Hadronisation Model (SHM) \cite{RefSHMNuclei} and the coalescence model \cite{RefCoalescence}. New observables such as (anti)nuclei event-by-event fluctuations, which are characterised by a high discriminating power between these two models, can help to shed light on the nuclear formation mechanisms \cite{RefCoalescenceFluctuations}. Moreover, among all species produced in collisions, light nuclei are the most sensitive to antiparticle-particle balance, which is encoded by the baryon chemical potential $\mu_B$. As a consequence, they provide a precise constraint on the value of $\mu_B$.

\section{Antideuteron event-by-event fluctuations}
\label{sec-1}
The properties of the (anti)nuclei multiplicity distributions are directly connected to the assumed (anti)nucleosynthesis mechanism  \cite{RefCoalescenceFluctuations}: in particular, a Poisson distribution is predicted by the Grand Canonical (GC) SHM \cite{RefBaseline}. Conservation laws and correlations among the particles produced in the collision can cause deviations from the Poisson baseline. This is the case for the canonical SHM and the coalescence model. The event-by-event fluctuations can be experimentally characterised by the $m$-th order cumulant $\kappa_m$: in this study, the ratio between the cumulants $\kappa_2$ and $\kappa_1$, which represent the variance and the average of the distribution respectively, is considered. Two versions of the coalescence model are taken into account: in model A, the multiplicities of the coalescing (anti)protons and (anti)neutrons are assumed to be correlated, while according to model B, (anti)protons and (anti)neutrons are emitted by independent sources.

\begin{figure}[h]
\centering
\subfloat{
    \centering
    \includegraphics[width=5.6cm,clip]{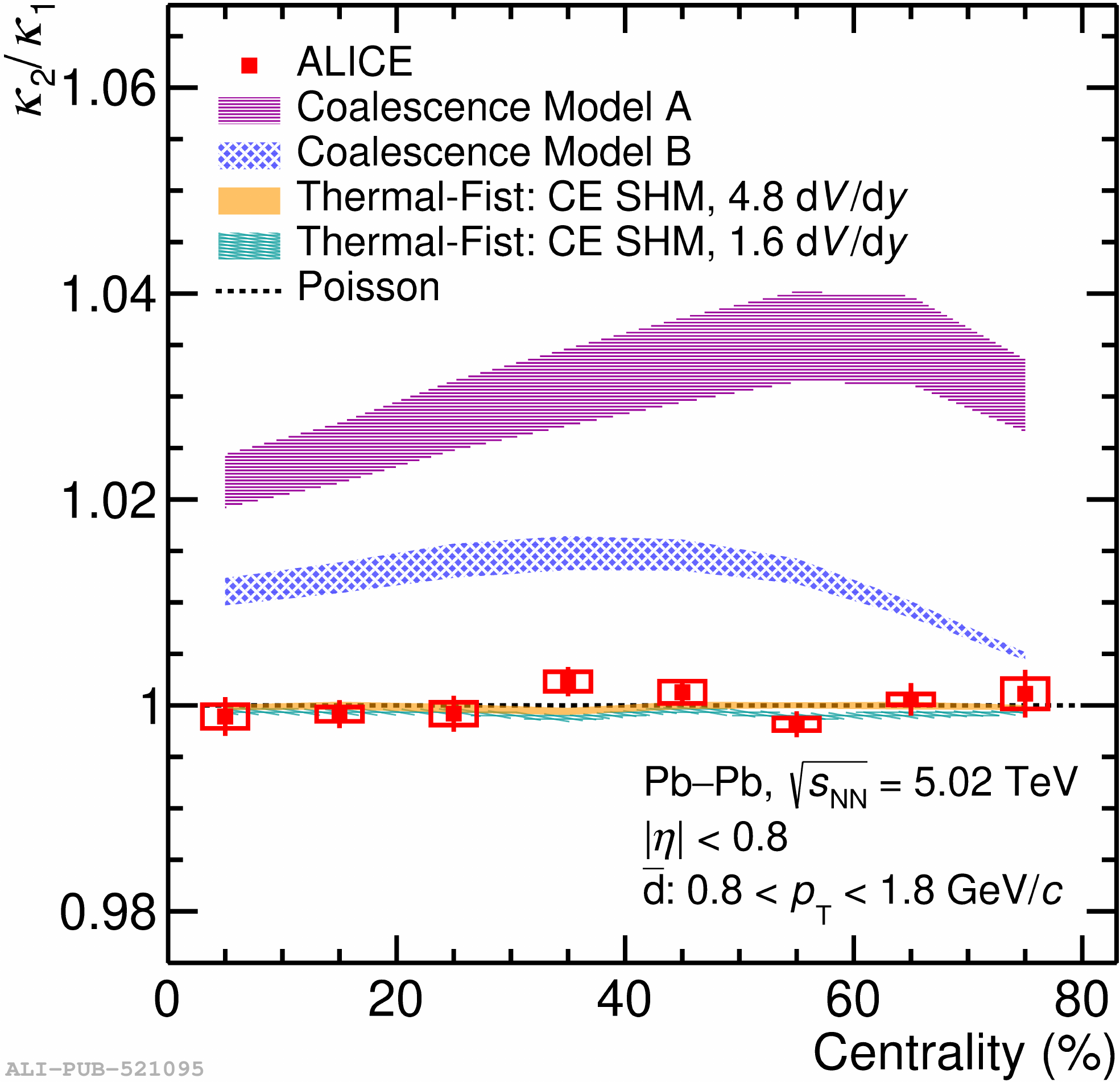}
}
\subfloat{
    \centering
    \includegraphics[width=5.6cm,clip]{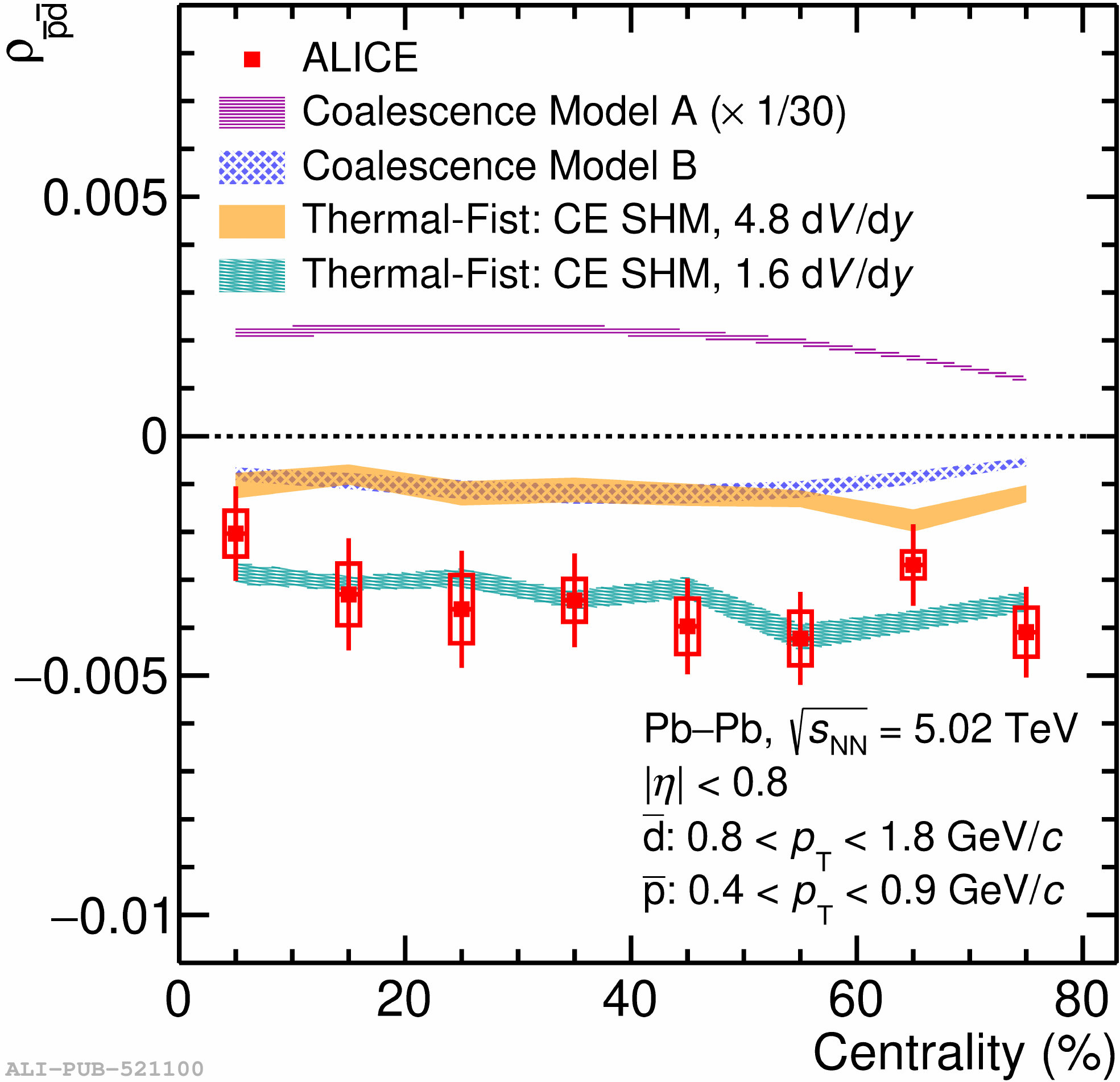}
}
\caption{(Left) Ratio of cumulants $\kappa_2$ and $\kappa_1$ of the antideuteron multiplicity distribution as a function of the event centrality. (Right) Antideuteron-antiproton correlation $\rho_{\overline{\mathrm{p}}\overline{\mathrm{d}}}$ as a function of the event centrality. In both plots, the ALICE measurements are shown as red points, with statistical and systematic uncertainties represented as vertical lines and boxes, respectively. The predictions obtained from both the coalescence and the statistical model are shown as coloured bands \cite{RefAntideuteronFluctuations}.}
\label{fig-1}       
\end{figure}

The $\kappa_2/\kappa_1$ ratio of antideuterons measured by the ALICE Collaboration in Pb--Pb collisions at $\sqrt{s_{\mathrm{NN}}}=5.02\ \mathrm{TeV}$ is shown in the left panel of figure~\ref{fig-1} as a function of the event centrality \cite{RefAntideuteronFluctuations}. Antideuterons are measured in a narrow transverse momentum ($p_\mathrm{T}$) range, $0.8<p_{\mathrm{T}}<1.8\ \mathrm{GeV}/c$, where the outstanding particle identification (PID) capabilities of ALICE can be fully exploited to select a high-purity antideuteron sample \cite{RefDeuterons}. Moreover, only the antimatter component is considered for this measurement as its observed yield is not contaminated by the products of spallation interactions of primary particles in the detector. Predictions of coalescence models A (violet) and B (blue), and those of the Canonical Ensemble (CE) SHM with baryon-number correlation volume $V_C=4.8\ \mathrm{d}V/\mathrm{d}y$ (yellow) and $V_C=1.6\ \mathrm{d}V/\mathrm{d}y$ (green) are each compared with the measured points in the left panel of figure~\ref{fig-1}. It can be observed that the measured $\kappa_2/\kappa_1$ ratio is consistent with unity, i.e. with the Poissonian limit, across centrality: the CE SHM is able to reproduce this result, as antideuterons constitute only a small fraction of the antibaryons produced in the collision, while both version A and B of the coalescence model overpredict the observed ratio.

Another variable considered in this study is the antiproton-antideuteron correlation, $\rho_{\overline{\mathrm{p}}\overline{\mathrm{d}}}$: the results obtained by the ALICE Collaboration as a function of the event centrality are shown in the right panel of figure~\ref{fig-1}. An anticorrelation of about $-0.5\%$ is observed across centrality: this result can be succesfully described by the CE SHM with $V_C=1.6\ \mathrm{d}V/\mathrm{d}y$, i.e. with a baryon-number correlation volume smaller than the one obtained in studies of net-proton fluctuations \cite{RefNetProtonFluctuations}. The observed anticorrelation can be also qualitatively described by the coalescence model B, while model A is ruled out by the comparison with data, as it predicts a large positive correlation of about 5\% between antiprotons and antideuterons. The CE SHM predictions are also in good agreement with $\rho_{\overline{\mathrm{p}}\overline{\mathrm{d}}}$  as a function of the acceptance window used for the measurement \cite{RefAntideuteronFluctuations}.


\section{The baryon chemical potential}
\label{sec-2}
As shown in section~\ref{sec-1}, (anti)nucleus production can be successfully described by statistical models: consequently, light (anti)nuclei can be used in turn to determine the model parameters, specifically the baryon chemical potential $\mu_B$. The $\mu_B$ measurement described in the following is based on the precise evaluation of antiparticle-to-particle ratios of different species ($\overline{h}/h$), which are connected to the value of $\mu_B$ via a relation obtained from the GC SHM under the assumption of strangeness neutrality \cite{RefRatiosSHM, RefBaryonStrangenessChemicalPotentials}:

\begin{equation}\label{eq-SHM-ratios}
    \overline{h}/h\propto\exp\left[-2\left(B+\frac{S}{3}\right)\frac{\mu_B}{T}-2I_3\frac{\mu_{I_3}}{T}\right]
\end{equation}

where $B$, $S$ and $I_3$ are the baryon number, strangeness and isospin third component of the species $h$, respectively, while $\mu_{I_3}$ is the isospin chemical potential. From equation~\ref{eq-SHM-ratios} it can be seen that the species characterised by a large baryon content are more sensitive to $\mu_B$; moreover, there is also a dependence on strangeness. The $\mu_B$ measurement presented here is based on the antiparticle-to-particle ratios of species reconstructed by ALICE that carry the highest values of $B$, i.e. protons, $^3\rm{He}$ and hypertriton ($^3_\Lambda\rm{H}$). The ratio of charged pion yields is also taken into account to put a tight constraint on $\mu_{I_3}$. In order to extract $\mu_B$, the chemical freeze-out temperature $T$ can be fixed to the value $T=156.2\pm1.5\ \rm{MeV}$ obtained from previous studies \cite{RefDecodingQCDNature}, since the dependence of ratios on $T$ in equation~\ref{eq-SHM-ratios} is negligible for $\mu_B\sim O(1\ \mathrm{MeV})$.

The data sample analysed consists of about $3\times 10^8$ Pb--Pb collisions at $\sqrt{s_{\rm{NN}}}=5.02\ \rm{TeV}$ recorded during the 2018 LHC run: the large sample size allows us to perform a centrality-differential study in three centrality classes, namely 0-5\%, 5-10\% and 30-50\%. The analysis of pions, protons and helium relies on standard rectangular selections, while the hypertriton measurement is based on machine learning selections \cite{ReFALICEHypertritonpPb}. For all the analysed particles, the contributions due to spallation and weak decays are subtracted to account only for the primary particle yields in the final results.

\begin{figure}
\centering
\includegraphics[width=12.5cm,clip]{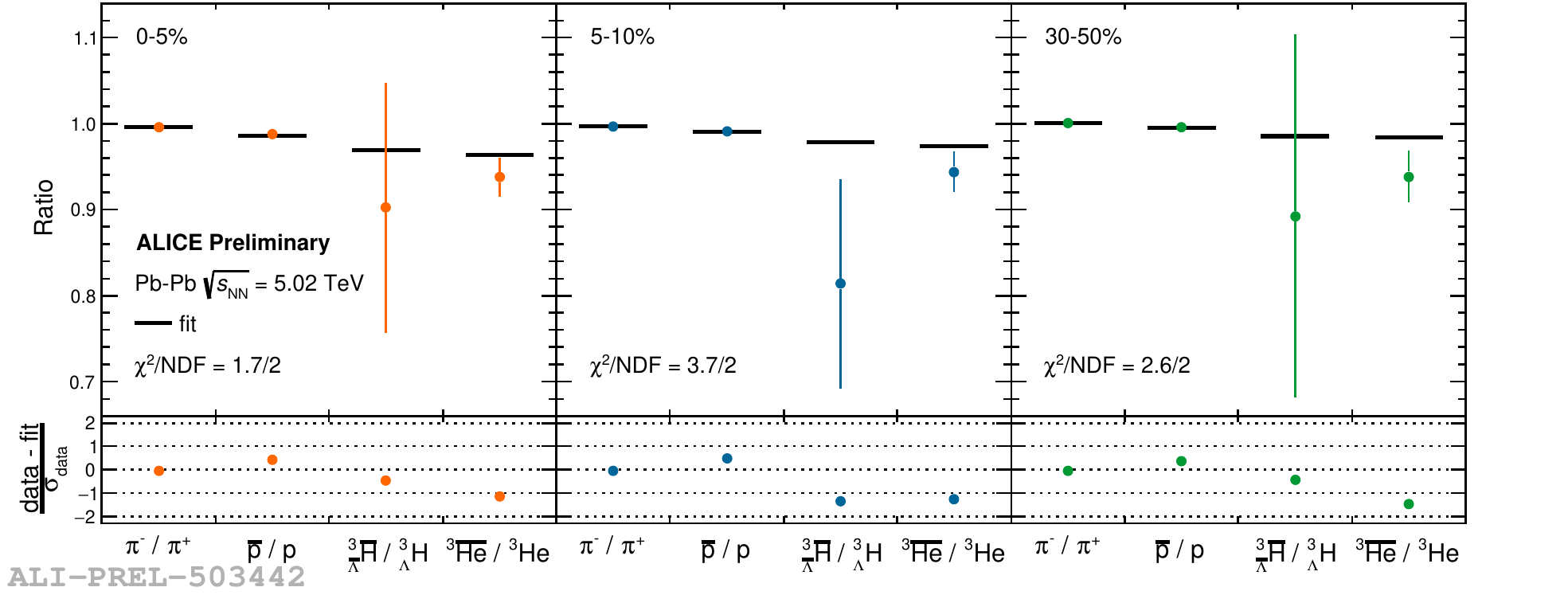}
\caption{Results of the SHM fit (black lines) to the measured antiparticle-to-particle ratios in the 0-5\% (orange), 5-10\% (blue) and 30-50\% (green) centrality classes. The vertical lines show the sum in quadrature of statistical and $p_{\mathrm{T}}$- and centrality-uncorrelated systematic uncertainties. The difference between the fit results and the measured ratios normalised to the total uncertainty is shown in the lower panel.}
\label{fig-2}       
\end{figure}

\begin{figure}
\centering
\sidecaption
\includegraphics[width=7.5cm,clip]{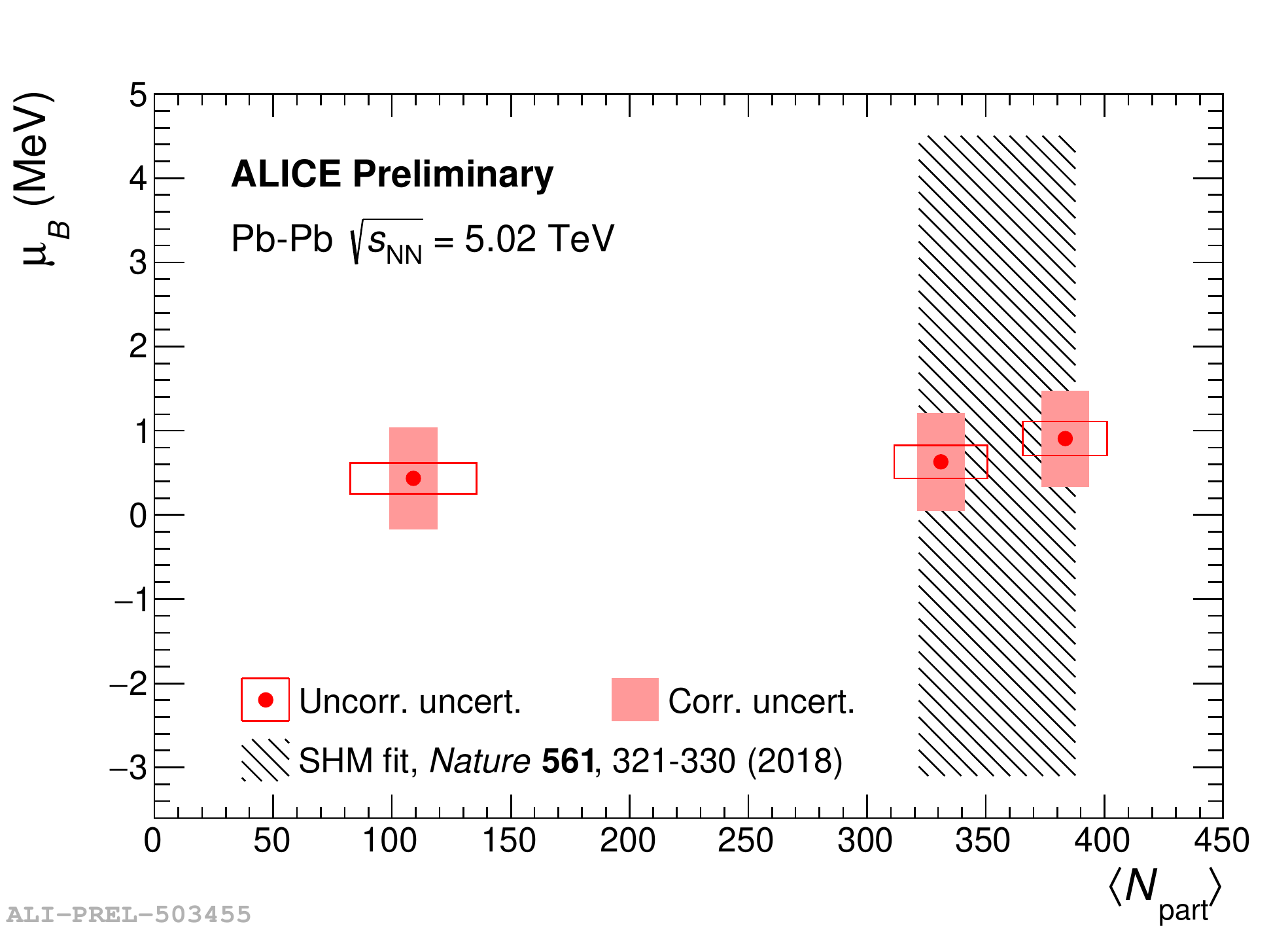}
\caption{Baryon chemical potential $\mu_B$ as a function of the average number of participating nucleons $\langle N_{\mathrm{part}}\rangle$. The uncorrelated uncertainties are shown as open boxes, while the correlated ones as coloured rectangles. The latest published measurement of the baryon chemical potential at the LHC, $\mu_B=0.7\pm 3.8\ \mathrm{MeV}$ \cite{RefDecodingQCDNature}, is represented as a black shaded area.}
\label{fig-3}       
\end{figure}

The ratios obtained for the three analysed centrality classes are shown in figure~\ref{fig-2}. A fit based on equation~\ref{eq-SHM-ratios} with $\mu_B$ and $\mu_{I_3}$ as free parameters is performed separately in each centrality interval: it can be observed that the baryon-number hierarchy predicted by the GC SHM is also present in the measured points. The $\mu_B$ values obtained are then shown as a function of the average number of participating nucleons $\langle N_{\rm{part}}\rangle$ in figure~\ref{fig-3}: the $\mu_B$ results obtained for the most central collisions are compatible with previous studies based on SHM fits of antiparticle and particle yields of different species \cite{RefDecodingQCDNature}, while showing an improvement in precision by about one order of magnitude; in addition, a possible centrality-dependence due to different baryon stopping at different centralities is excluded with the current precision.

\section{Conclusions}
The study of antideuteron event-by-event fluctuations offers a powerful probe of (anti)nucleosynthesis: the results shown in this contribution favour the CE SHM predictions over the coalescence ones. The large data samples that will be collected during LHC Runs 3 and 4 will allow us to perform more precise measurements of this observable. In addition, the most precise $\mu_B$ measurement ever performed is reported here: this study will be further extended by testing the strangeness and isospin dependence of ratios via the strange baryons $\Lambda$ and $\Omega$, and triton.

%

\begin{thebibliography}{}
%
%

\bibitem{RefSHMNuclei}
A. Andronic, P. Braun-Munzinger, J. Stachel, H. St\"ocker, Phys. Lett. B \textbf{697}, 203--207 (2011)
\bibitem{RefCoalescence}
S.T. Butler, C.A. Pearson, Phys. Rev. \textbf{129}, 836--842 (1963)
\bibitem{RefCoalescenceFluctuations}
Z. Feckov\'a, J. Steinheimer, B. Tom\'a\ifmmode \check{s}\else \v{s}\fi{}ik, M. Bleicher, Phys. Rev. C \textbf{93}, 054906 (2016)
\bibitem{RefBaseline}
P. Braun-Munzinger, B. Friman, K. Redlich, A. Rustamov, J. Stachel, arXiv:2007.02463 [nucl-th] (2020)
\bibitem{RefAntideuteronFluctuations}
ALICE Collaboration, arXiv:2204.10166 [nucl-ex] (2022)
\bibitem{RefDeuterons}
J. Adam, et. al (ALICE Collaboration), Phys. Rev. C \textbf{93}, 024917 (2015)
\bibitem{RefNetProtonFluctuations}
S. Acharya, et al. (ALICE Collaboration), Phys. Lett. B \textbf{807}, 135564 (2020)
\bibitem{RefRatiosSHM}
J. Cleymans, I. Kraus, H. Oeschler, K. Redlich, S. Wheaton, Phys. Rev. C \textbf{74}, 034903 (2006)
\bibitem{RefBaryonStrangenessChemicalPotentials}
J. Cleymans, H. Satz, Z. Phys. C \textbf{57}, 135--147 (1993)
\bibitem{RefDecodingQCDNature}
A. Andronic, P. Braun-Munzinger, K. Redlich, J. Stachel, Nature \textbf{561}, 321--330 (2018)
\bibitem{ReFALICEHypertritonpPb}
S. Acharya, et. al (ALICE Collaboration), arXiv:2107.10627v1 [nucl-ex] (2021)

\end{thebibliography}
%
%

\end{document}